\documentclass[11pt]{article}
\usepackage{amsfonts,amsbsy,amsmath,amssymb}

\newcommand{\BE}{\begin{equation}}
\newcommand{\EE}{\end{equation}}
\newcommand{\BA}{\begin{eqnarray}}
\newcommand{\EA}{\end{eqnarray}}
\newcommand{\Pp}{{\scriptstyle{{\rm P}}}}

\newcommand{\msbar}{\overline{\rm MS}}

\begin{document}

\begin{titlepage}

\vspace*{22mm}
\begin{center}
              {\LARGE\bf  `Maximal conformality' does not work}
\vspace{23mm}\\
{\large P. M. Stevenson}
\vspace{4mm}\\
{\it
T.W. Bonner Laboratory, Department of Physics and Astronomy,\\
Rice University, Houston, TX 77251, USA}

\vspace{30mm}

{\bf Abstract:}

\end{center}

\begin{quote}
The so-called ``principle of maximal conformality'' is ineffective and does nothing to resolve the 
renormalization-scheme-dependence problem.  Some essential facts about that problem are summarized.  
It is stressed that RG invariance is a symmetry and that any viable method for resolving the scheme-dependence 
problem should  be formulatable in terms of the invariants of that symmetry.  
\end{quote}

\end{titlepage}

\setcounter{equation}{0}

    This paper is provoked by Ref. \cite{gbrww}, the latest in a long line of papers by Brodsky and collaborators 
going back to the BLM paper \cite{BLM} of 1983.\footnote{
      There are many other papers by Brodsky with other collaborators; see references in \cite{gbrww}.  
(I believe that Lepage and Mackenzie were not co-authors on any of these subsequent papers.)  
}
I will  state my opinion bluntly:  All of this work is misguided.   It is confused about basic issues and often asserts or 
implies that results are renormalization-scheme (RS) invariant when they are not.  I shall not attempt to justify my 
opinion with a point-by-point critique.\footnote{
     The basic deficiency of BLM was pointed out politely in Ref.~\cite{CelSte}.  The same point was later made more 
forcefully by Ch\'{y}la \cite{chyla}.
 }
 Rather, I aim to explain, concisely, what the RS-dependence problem is.  

     Where Brodsky {\it et al} are entirely correct is in forthrightly criticizing the shameful inadequacy of current practice 
in QCD phenomenology when faced with the RS-dependence problem.  The nature of this problem seems to be widely 
misunderstood.  (Frequent mis-characterizations of the issue as ``scale fixing'' are symptoms of this misunderstanding.)  
Since 1981 \cite{OPT} I have consistently advocated a method to resolve the problem and my arguments are set out in full 
in a recent book \cite{PMSbook}.  Here I set advocacy aside (until the very end).   I shall try to differentiate carefully between 
matters of opinion and matters of fact.  As I shall explain, one can objectively distinguish, at a very basic level, between 
{\it methods that work} and {\it methods that don't work.}

       Renormalization-Group (RG) invariance\footnote{
      The context here is an asymptotically-free renormalizable quantum field theory with a single coupling; in particular QCD 
with $n_f$ species of massless fermions.  (For simplicity I shall ignore the issue of `matching' such theories, each with a fixed 
$n_f$, to real QCD where the known quarks have a certain pattern of masses.)  In other contexts the term {\it Renormalization Group} is 
employed to cover a wide array of diverse and powerful ideas.  Without wishing to downplay the importance of such ideas, I just need to 
emphasize that in the present, limited context the RG is what Stueckelberg and Peterman \cite{SP} originally envisaged; an exact symmetry of 
a renormalizable theory.  Note that the cutoff plays no role here; any regularization has been and gone, having done its job.  
}
is a symmetry.  The renormalized coupling constant (couplant) 
$a \equiv \alpha_s/\pi \equiv g^2/(4 \pi^2)$ has no unique definition and can be transformed to another one, $a'$, by 
transformations of the form
\BE
\label{ap}
a' = a(1+ v_1 a + v_2 a^2 + \ldots),
\EE
where the $v_i$ are finite but otherwise arbitrary.  The coefficients in the perturbation series of a physical 
observable\footnote{
       The leading-order coefficient (which is RS invariant) has been scaled out for convenience.  More generally ${\cal R}$ could begin 
with $a^\Pp(1+ \ldots)$, but the generalization is entirely straightforward and brings in no new features so, for simplicity, I consider 
here only $\Pp=1$.  
}
\BE
{\cal R} = a(1 +  r_1 a + r_2 a^2 + \ldots), 
\EE
then transform as 
\BA
{r_1}^\prime & = & r_1-v_1, \nonumber \\
{r_2}^\prime & = & r_2-2 r_1 v_1 + 2 v_1^2-v_2, \label{rp}  \\
& & \mbox{\rm etc.,} \nonumber
\EA
so that ${\cal R}$ remains invariant.  
This is an exact symmetry of the theory respected, formally, by perturbation theory.  The RS-dependence problem 
arises only when the perturbation series is {\it truncated}.  Because the problem arises solely from the approximation 
its solution is to be found (in my opinion) not by examining the peculiarities of any specific theory, but by thinking more clearly 
about approximations.  

       Approximations always unavoidably involve some uncertainty and ambiguity.  One has always to decide 
`What are we calculating?' and `What is the form of the approximant?'  For ordinary power series the standard 
form of approximant is the partial sum, a truncated series.  (Here one must truncate not only the series for ${\cal R}$ 
but also that for the $\beta$ function and, since the order of the error is controlled by whichever truncation is the more 
severe, it is natural to use the same number of terms in each series.)  For an ordinary power series we would then have 
a definite result.   Here, however, we are faced with a major, open-ended ambiguity because all the series coefficients 
are RS dependent and the cancellations that preserve the all-orders result are incomplete, being spoiled by the truncations.  

     Any renormalization procedure (definition of an ``$a$'') introduces a renormalization scale $\mu$.  The dependence 
of $a$ on $\mu$ is given by the $\beta$ function:
\BE
 \mu \frac{d a}{d \mu} \equiv \beta(a) = -b a^2(1+ c a + c_2 a^2 + \ldots).  
\EE
A change in the value of $\mu$ is just a special case of a change of RS.  The coefficients $r_i$ can be changed by a scale 
transformation $\mu \to \mu'$, but they can also be changed by other choices that one may call the `renormalization 
prescription' so that scheme dependence is a combination of scale and prescription dependence.  Any method that 
fixes $\mu$ without fixing the prescription (or {\it vice versa}) achieves precisely {\it nothing.}  That is the key reason why 
BLM and its descendant methods do not work.  

      Note that even the next-to-leading order (NLO) coefficient $r_1$ depends on both scale and prescription.  A seeming puzzle -- 
to be explained later -- is that there are not two degrees of freedom but only one, embodied in the single arbitrary 
parameter $v_1$ at NLO.  

     I now state some facts about RS dependence.  Some are very well known but others seem to have been forgotten 
or overlooked.  

    (1) The first two coefficients $b$ and $c$ of the $\beta$ function are RS invariant, while $c_2, c_3, \ldots$ are not
(for instance, ${c_2}^\prime = c_2 + v_2 - v_1^2-v_1 c$)  because $ \beta'(a') \equiv \mu \frac{d a'}{d \mu} = \frac{d a'}{d a} \beta(a)$ 
\cite{tH}.

    (2) Integrating the $\beta$-function equation requires a constant of integration which brings in a scale parameter 
$\tilde{\Lambda}$ that can be conveniently defined by 
\BE
\label{intbeta0}
\ln(\mu/{\tilde{\Lambda}}) =
\lim_{\delta \to 0} \left( \int_\delta^{a} 
\frac{d x}{ \beta(x)} + {\cal C(\delta)} \right), 
\EE
with
\BE
\label{Cdeltadef}
{\cal C}(\delta) = \int_\delta^{\infty} 
\frac{d x}{ b x^2(1 + c x)}.  
\EE
Note that ${\cal C}(\delta)$ involves only $b$ and $c$ and so is RS invariant.  

   (3)  The $\tilde{\Lambda}$ parameter is RS dependent, but in a simple and definite way given by the Celmaster-Gonsalves 
(CG) relation \cite{CG}.  This states that if two prescriptions (schemes with the same value of $\mu$) are related by Eq.~(\ref{ap}) 
then 
\BE
\label{CG0a}
\ln(\tilde{\Lambda}^{\prime}/\tilde{\Lambda}) = v_1/b.
\EE 
This result is {\it exact} and does not involve the $v_2, v_3, \ldots$ coefficients. A proof of this relation is given in Appendix A.  
The single free parameter of QCD, in my opinion, can best be taken to be the $\tilde{\Lambda}$ parameter of some specified 
reference prescription.\footnote{
      While the choice of `reference prescription'  and the definition of $\tilde{\Lambda}$ (involving a specific choice of ${\cal C}(\delta)$) 
are arbitrary conventions, the conversion between any two choices can be made exactly, with no ``${}+\ldots$'' ambiguities.  
}

     (4)  Physical quantities (unlike Green's functions and other theoretical entities) are RS invariant.  That fact can be expressed 
symbolically as 
\BE
\label{rg} 
 0 = \frac{d{\cal R}}{d(RS)} =  
\left. \frac{\partial {\cal R}}{\partial (RS)} \right|_a  +
\frac{da}{d(RS)} \frac{\partial {\cal R}}{\partial a},
\EE
where the first term represents RS dependence from the series coefficients, $r_i$ (differentiation pretending that $a$ is constant), 
and the second represents the compensating RS dependence from the couplant, $a$.  The familiar RG equation expressing the 
$\mu$ independence of ${\cal R}$ is just a special case.  

     (5)  At NLO, as noted above, the coefficient $r_1$ depends on both $\mu$ and the prescription.  The resolution of the puzzle mentioned 
earlier is that $r_1$ depends on these two things only through a {\it single} variable, $\mu/\tilde{\Lambda}$.  There is an invariant quantity
\BE
\label{rho1Q}
{\boldsymbol \rho}_1(Q) \equiv b \ln(\mu/\tilde{\Lambda}) - r_1.  
\EE
The calculated $r_1$ coefficient, as is well known, always has the form 
\BE
\label{r1form}
r_1= b \ln(\mu/Q) + r_{1,o},
\EE
where $Q$ is some kinematic variable of the ${\cal R}$ quantity.  Thus the $\mu$ dependence cancels in ${\boldsymbol \rho}_1(Q)$, 
but equally importantly so does the prescription dependence, since that of $ r_{1,o}$ is exactly cancelled by that of $\tilde{\Lambda}$, 
thanks to the CG relation.  Also note that one may change one's mind about what ``$Q$'' is, since that is just moving a 
$b \ln(Q_{\rm new}/Q)$ term between the two terms of Eq.~(\ref{r1form}).  

   (6)  At higher orders the other RS variables, besides $\mu/\tilde{\Lambda}$, can be taken to be  the $\beta$-function coefficients 
$c_2,c_3,\ldots$.\footnote{
      The only RS-dependent variables in Eq.~(\ref{intbeta0}) are $\mu/\tilde{\Lambda}$ and $c_2, c_3, \ldots$, so $a$ can depend 
on RS only through those variables. 
}
The symbolic RG equation (\ref{rg}) can then be expressed concretely as a set of equations\footnote{
     Plus, of course, the usual RG equation for the $\mu$ (or rather the $\mu/\tilde{\Lambda}$) dependence, which is naturally the 
``j=1'' counterpart of these other equations \cite{PMSbook}.
}
\BE
\frac{\partial {\cal R}}{\partial c_j} = 
\left( \left. \frac{\partial}{\partial c_j} \right|_a + 
\beta_j(a) \frac{\partial}{\partial a} \right) {\cal R} = 0, 
\quad\quad j = 2,3,\ldots,  
\EE
involving some new functions $\beta_j(a)$ defined by 
\BE
\beta_j(a) \equiv \frac{\partial a}{\partial c_j}.
\EE
These $\beta_j(a)$ functions are fixed in terms of the $\beta(a)$ function \cite{OPT,PMSbook}.  

   (7)  There are further  RS invariants (besides $b$, $c$ and ${\boldsymbol \rho}_1(Q)$) at higher orders.   The next is 
\BE
\label{rho2}
\rho_2 \equiv  c_2 + r_2 - c r_1 - r_1^2.  
\EE
(Its RS invariance may easily be checked by substituting for ${r_1}^\prime, {r_2}^\prime$ and ${c_2}^\prime$ and observing 
the cancellation of all $v_1$ and $v_2$ dependence.)  These $\rho_2, \rho_3, \ldots$ invariants are $Q$ independent, as a 
consequence of the cancellation of $\mu$ dependence.  

     None of the points above is in any way dependent upon Ref.~\cite{OPT}'s {\it Principle of Minimal Sensitivity} (PMS) criterion:  
These are simply {\it facts} about the structure of renormalized perturbation theory.  

     I can now be more precise about the distinction between methods that ``work'' and methods that ``don't work.''  A method that 
``works'' is one where the final result does not depend upon which RS is used initially for the Feynman-diagram calculations.   
Methods that ``work'' include PMS \cite{OPT,PMSbook}, FAC \cite{FAC}, and infinitely many others that one might dream up.  
Such methods may always be formulated as a two step process (i) calculate all of the invariants to the appropriate order in any 
convenient RS and then (ii) follow some recipe, involving only the invariants as input, to get the result.  The {\it motivation} for 
any proposed recipe is of course debatable --  but it ought to be unnecessary to discuss methods that simply do not work.  

     Let me try to clarify exactly what is wrong with Brodsky {\it et al}'s method(s).  Firstly, the idea that one should demand 
``maximal conformality'' is not credible:   One can, by RS choice, achieve {\it exact} conformality, as is shown in Appendix B.  
Secondly, their notion that some pieces of QCD are `conformal' while other pieces are not is untenable. 
They write the $r_1$ coefficient as
\BE
r_1 = A n_f + B,
\EE
and claim that
\BE
C_1^*= \frac{33}{2}A + B
\EE
is the `conformal' coefficient, while the other piece of $r_1$, namely $-3 A b$ (where $b=(33-2 n_f)/6$) is the `non-conformal' piece.     
However, making $b$ vanish (by setting $n_f=33/2$) does not produce a conformal theory (the $\beta$ function does 
not vanish but becomes of the form $h a^3(1+ \ldots)$, where $h=- b c=- (153-19 n_f)/12 =107/8$).\footnote{
 One then has a non-asymptotically free `delicate' theory in which no divergences appear at one loop, but only at two loops 
and beyond.     There is no analogy with QED, where $b$ is proportional to $n_f$.  For $n_f=0$ the whole QED $\beta$ 
function vanishes and the theory becomes conformal -- but that is because it becomes a free field theory of  massless 
photons.  There are no divergences (excepting the unobservable vacuum energy density).  QCD with $n_f=33/2$ is quite different.  
} 

     The really key point, though, is that $C_1^*$ is not RS invariant; it can be changed arbitrarily by a scheme transformation.  
If one takes $v_1$ in Eq.~(\ref{ap}) to be of the form\footnote{
    In fact there is no good reason why $v_1$ should be restricted to being linear in $n_f$.  However, I waive that objection 
for the sake of argument.  
}
\BE
v_1=v_{10} n_f + v_{11}
\EE
then $C_1^*$ changes by a term $-(\frac{33}{2}v_{10}+v_{11})$, which could have any value.  Now it is true that for scale 
transformations (or for the specific relation between the minimal subtraction (MS) and $\msbar$ prescriptions) the $v_1$ is proportional 
to $b$ and the change in $C_1^*$ vanishes, but these are very special cases.  It is quite natural, in Feynman-diagram terms, for $v_{10}$ 
to be changed independently of $v_{11}$.  While in  $\msbar$ the trace of the unit Dirac matrix in $d$ dimensions is always set to $4$, one may 
well choose a different convention (perhaps to interpolate the formula $2^{d/2}$ appropriate for even integer dimensions).  That would affect 
only the fermion-loop contributions proportional to $n_f$.\footnote{
    While $n_f$ dependence has no real relevance to the RS-dependence problem, it is not without interest.  In particular there are
large-$b$ and small-$b$ approximations to QCD.  These approximations are, however, quite distinct from renormalized perturbation theory.  
The small-$b$ case (the Banks-Zaks expansion) is a topic that I have written about elsewhere (see \cite{PMSbook} and references 
therein).  For a large class of important physical quantities the invariants $\rho_j$ can be decomposed in powers of $n_f$, and hence 
in powers of $b$.  However, the ${\boldsymbol \rho}_1(Q)$ invariant cannot be so decomposed because it is not really meaningful to ask 
how  $\tilde{\Lambda}$ depends on $n_f$; after all, theories with different $n_f$'s are just different theories and describe different model 
universes.  None of this has anything to do with the practical question of how best to approximate real QCD, with its particular quark masses, 
by a series of massless theories with different numbers of `active' flavours.  That is a whole other kind of approximation, and involves deciding 
where to put the `thresholds' and how to match the $\tilde{\Lambda}$ parameters.  For my views on those issues see Ref.~\cite{lowen}.
}

      Ref. \cite{gbrww} says (below Eq. (3.5)) that $C_1^*$ ``is the {\it conformal} coefficient, i.e., the NLO coefficient not depending 
on the RS and scale.''  That is untrue, as has just been shown.  The fact that $C_1^*$ is prescription dependent was explicitly admitted 
in the BLM paper \cite{BLM}.\footnote{
     See the sentences after Eq. (5) in \cite{BLM}.   Celmaster and I had written to BLM in late 1982 pointing out the 
non-invariance of $C_1^*$. See \cite{CelSte}.  
}
An initial choice of renormalization prescription actually has to be specified and all results depend 
upon that arbitrary choice.  Later papers, in their titles, abstracts and introductions, often imply RS-invariant results -- but, in the small print, it turns out 
that the results are only ``approximately'' RS invariant, with long discussions of how to deal with the ``residual scheme dependence'' (as if that wasn't 
the whole problem from the beginning!).  

     The reader might wonder why, in purely phenomenological terms, Brodsky {\it et al}'s final numerical results often seem quite good.  
The reason is interesting.  Because they must eventually admit that their method leaves ``residual scheme dependence'' they go to some pains, 
by various means, to ensure that this residual scheme dependence is numerically fairly insignificant (for `small' changes of prescription or scale) in 
comparison with the estimated error of the approximation.  In doing so they are ensuring that they are not too far from the minimal-sensitivity point 
where any small scheme change produces no first-order change in the approximant's value.  But why do that  in a such a laborious, {\it ad hoc}, 
back-door fashion when one can get to the minimal-sensitivity point directly and systematically?   

     Many of the facts outlined above are explained in more detail in Ref.~\cite{PMSbook}.  The treatment of cases 
involving factorization-scheme dependence is also discussed there.

\newpage

\section*{Appendix A:  Proof of the CG relation}

      There is nothing wrong with the original proof given by Celmaster and Gonsalves \cite{CG} but the following proof (a slight variant of one due to 
Osborn \cite{osborn,PMSbook}) offers more insight into why the result is {\it exact}.  Consider two prescriptions (schemes with the same value of $\mu$).  
The counterpart to Eq.~(\ref{intbeta0}) in the primed scheme is 
\BE
\label{intbetapr}
\ln(\mu/{\tilde{\Lambda}}') =
\lim_{\delta' \to 0} \left( \int_{\delta'}^{a'} 
\frac{d x'}{ \beta'(x')} + {\cal C(\delta')} \right).  
\EE
The -- seemingly redundant -- inclusion of primes on the dummy integration variable $x$ and the $\delta$ parameter is actually convenient.  
One can now make a change of variables 
\BE
x' = x(1+ v_1 x + v_2 x^2 + \ldots),
\EE
(and similarly for $\delta'$) with the same $v_i$ coefficients as for $a'$.  Thus, when one subtracts Eq.~(\ref{intbetapr}) from 
Eq.~(\ref{intbeta0}) one finds that the integral terms cancel {\it exactly} because the $\beta$ function transforms as 
$\beta'(x')=(d x'/dx) \beta(x)$.   That leaves one with 
\BE
\ln(\mu/{\tilde{\Lambda}}) - \ln(\mu/{\tilde{\Lambda}}')  = 
\lim_{\delta \to 0} \left({\cal C}(\delta) -{\cal C}(\delta') \right)
\EE
The integral ${\cal C}(\delta)$ of Eq.~(\ref{Cdeltadef}) is easily evaluated and one finds that 
\BE
{\cal C}(\delta) -{\cal C}(\delta') =  \frac{1}{b} \left( \frac{1}{\delta}-\frac{1}{\delta'} + O(\delta) \right) 
\EE
and so, taking the limit $\delta \to 0$, one has 
\BE
\ln({\tilde{\Lambda}}'/{\tilde{\Lambda}}) = v_1/b.
\EE

\section*{Appendix B:  
Achieving maximal conformality}

    If ``maximal conformality'' is what we desire then we may use the RS choice to achieve {\it exact} conformality, with our result 
for ${\cal R}$ being energy independent.   There are several ways to do this.  One is to adjust the scheme, decreasing the $\mu/\tilde{\Lambda}$ 
value, until the coefficient $r_1$ becomes so large and negative that $r_1 a=-1$.  The NLO result is then ${\cal R}=a(1-1)=0$.  Alternatively, we may 
make $\mu/\tilde{\Lambda}$ arbitrarily large, and hence $a$ arbitrarily small; the $r_1$ coefficient then becomes large and positive, approaching 
$1/a$ as $a \to 0$;  our NLO result is then $a(1+1)=2 a \to 0$.  We may easily extend these stratagems to higher orders 
(no actual Feynman-diagram calculations are needed!) to achieve ${\cal R}=0$ to any order.  To be clear; I am not advocating such RS's; 
this is satire, a {\it reductio ad absurdum}. The serious point is this:  Whatever it is that characterizes a ``good'' choice of RS, 
``maximal conformality'' is certainly not it.

\end{document}